\documentclass[aps,amsfonts,amsmath,prd,preprint,tightenlines,nofootinbib,showpacs]{revtex4}

\def\be{\begin{equation}}
\def\ee{\end{equation}}

\newtheorem{theorem}{Theorem}
\newtheorem{condition}{Condition}
\newtheorem{lemma}{Lemma}

\def\scri{{\cal I}}

\begin{document}

\title{Achronal averaged null energy condition}
\author{Noah Graham}
\email{ngraham@middlebury.edu}
\affiliation{Department of Physics,
Middlebury College, Middlebury, VT  05753}
\affiliation{Center for Theoretical Physics, Laboratory for
Nuclear Science, and Department of Physics,
Massachusetts Institute of Technology,
Cambridge, MA 02139}

\author{Ken D. Olum}
\email{kdo@cosmos.phy.tufts.edu}
\affiliation{Institute of Cosmology, Department of Physics and
Astronomy, Tufts University, Medford, MA  02155}

\begin{abstract}
The averaged null energy condition (ANEC) requires that the integral
over a complete null geodesic of the stress-energy tensor projected
onto the geodesic tangent vector is never negative.  This condition is
sufficient to prove many important theorems in general relativity, but
it is violated by quantum fields in curved spacetime.  However there
is a weaker condition, which is free of known violations, requiring
only that there is no self-consistent spacetime in semiclassical
gravity in which ANEC is violated on a complete, {\em achronal} null
geodesic.  We indicate why such a condition might be expected to hold
and show that it is sufficient to rule out closed timelike curves and
wormholes connecting different asymptotically flat regions.
\end{abstract}


\maketitle

\section{Introduction}
General relativity alone allows any smooth Lorentzian manifold to be a
spacetime.  Given a desired spacetime geometry, one simply solves
Einstein's equations in reverse to determine the stress-energy tensor
$T_{ab}$ needed to produce it.  Thus any restrictions on exotic phenomena,
such as wormholes or time machines, must be given in terms of \emph{energy
conditions} that restrict the set of possible stress-energy tensors.

One would hope that such conditions are satisfied in semiclassical
gravity, i.e., that every quantum state would satisfy a condition on
$\langle T_{ab}\rangle$, where the angle brackets denote the quantum
mechanical average.  Unfortunately, all the conditions usually
considered are known to be violated by quantum fields in in curved
spacetime.  The weakest such condition is the averaged null energy
condition (ANEC), which requires that
\be
\int_\gamma T_{ab} k^a k^b > 0
\ee
where the integral is taken over a complete null geodesic $\gamma$
with tangent vector $k^a$.  In flat space, this condition has been
found to be obeyed by quantum fields in many backgrounds where one
might expect it to be violated, such a domain wall
\cite{hep-th/0211244} or a Casimir plate with a hole
\cite{hep-th/0506136}.  The latter result has has been generalized to
arbitrary Casimir systems, as long as the geodesic does not intersect
or asymptotically approach the plates \cite{gr-qc/0609007}.  However,
a quantum scalar field in a spacetime compactified in one spatial
dimension or in a Schwarzschild spacetime around a black hole violates
ANEC \cite{gr-qc/9604008}.

We will therefore consider a weaker condition, which for clarity
we will call the self-consistent achronal
averaged null energy condition:
\begin{condition}[self-consistent achronal ANEC]
There is no self-consistent solution in semiclassical gravity in which
ANEC is violated on a complete, achronal null geodesic.
\end{condition}
We conjecture that all semiclassical systems obey self-consistent
achronal ANEC.

There are two changes here from the usual ANEC.  The first is that we
require it to hold only on achronal geodesics (those that do not
contain any points connected by a timelike path).  This requirement
avoids violation by compactified spacetimes and Schwarzschild
spacetimes, as discussed in Sec.\ \ref{sec:counterexamples}.  But for
proving theorems, this restriction is generally unimportant, as
discussed in Sec.\ \ref{sec:proofs}.  The null geodesics used in the
proofs are generally those which represent the ``fastest'' paths from
one place to another or those which are part of a horizon separating
two parts of spacetime that have different causal relationships to
some certain region.  To play either of these roles, geodesics must be
achronal.

The second change is that we are no longer discussing the
stress-energy tensor of the fluctuating quantum field separately from
the stress-energy tensor of the background.  Instead, we consider a
situation in which Einstein's equation relates the spacetime curvature
to the full stress-energy tensor, comprising both the classical
contribution from ordinary matter and the induced quantum contribution
one obtains in the background of this curvature.  This approach avoids
a potential violation due to the scale anomaly, as discussed in Sec.\
\ref{sec:anomaly}.

The idea of requiring ANEC to hold only on achronal geodesics appears
to have been introduced by Wald and Yurtserver \cite{PHRVA.D44.403},
who proved Condition 1 for a massless scalar field in $1+1$
dimensions.  In that case, however, all geodesics are achronal unless
the spacetime is periodic in space, in which case no geodesics are
achronal.  The idea of not requiring ANEC to hold on test fields but
only in a self-consistent system  appears to have been
introduced by Penrose, Sorkin and Woolgar \cite{gr-qc/9301015}.
Self-consistent systems were studied extensively by Flanagan and Wald
\cite{gr-qc/9602052}.

We restrict this analysis to semiclassical gravity, meaning that
Condition 1 should be expected to hold only in cases
where the curvature is well below the Planck scale, where a
semiclassical analysis of a quantum field on a classical curved space
background is applicable.  This condition eliminates classical 
violations of ANEC \cite{Barcelo:2000zf}, because they are
obtained by increasing the fields to Planck scale values.  This
process increases the effective gravitational coupling $G$ through a
region where it diverges, and thus clearly leaves the
semiclassical regime.

An immediate consequence of Condition 1 is the following:
\begin{lemma}
In a generic spacetime obeying Condition 1, there are no complete,
achronal null geodesics.
\end{lemma}
By a generic spacetime we mean one that obeys the null generic
condition, which states that every complete null geodesic contains a
point where $k^ak^bk_{[c} R_{d]ab[e} k_{f]}\ne0$, where $k^a$ is the
tangent vector.  This condition says that every null geodesic is
affected by some matter or tidal force.  In such a spacetime, every
complete null geodesic that obeys ANEC has a pair of conjugate points
\cite{Borde:1987qr} and thus is chronal \cite{H&E}.

Why should one believe that self-consistent achronal ANEC holds, when
other conditions have failed?  First of all, no violations are known,
as we discuss below.  But we also suggest that self-consistent
achronal ANEC can be proved along the lines of
Ref. \cite{gr-qc/0609007}.  That paper showed that ANEC holds for a
minimally coupled scalar field on a geodesic that travels in a tube of
flat space embedded in an arbitrary curved spacetime, assuming that
the causal structure of the tube is unaffected by the exterior
spacetime.  This last condition guarantees that the geodesic is
achronal.  We expect that any spacetime could be slightly deformed in
the vicinity of a given geodesic to produce the necessary tube, so
that self-consistent achronal ANEC could be proved along similar
lines, but such a proof will have to await future work.

\section{Explicit counterexamples to ANEC}

\label{sec:counterexamples}

To our knowledge, there are two specific spacetimes in which ANEC has
been explicitly calculated and found to be violated.  The first is
Minkowski space compactified in one spatial dimension.  For example,
one could identify the surfaces $z = 0$ and $z = L$.  The resulting
situation is very much analogous to the Casimir effect.  ANEC is
violated on any geodesic that does not remain at constant $z$.
However, no such geodesic is achronal.  Since the system is invariant
under all translations and under boosts in the $x$ and $y$ directions,
it suffices to consider the geodesic through the origin in the $z$
direction.  This returns infinitely often to $x = y = z = 0$, and thus
is chronal.

The second known violation is in Schwarzschild spacetime, in
particular in the Boulware vacuum state \cite{gr-qc/9604008}.  But
every complete geodesic in the Schwarzschild spacetime is
chronal,\footnote{The radial geodesic is achronal but not complete.  In
the Schwarzschild metric, $k^ak^bk_{[t} R_{r]ab[t} k_{r]} =
-(3M/r^3)\sin\alpha$, where $\alpha$ is the angle between the
direction of $k$ and the radial direction.  Thus the null generic
condition holds for any non-radial motion, so any complete geodesic
contains conjugate points and thus is chronal.} so self-consistent
achronal ANEC is (trivially) satisfied.

In addition, Flanagan and Wald \cite{gr-qc/9602052} found violations
of ANEC in self-consistent perturbation theory about Minkowski space.
Although they stated ANEC in the achronal form, they did not discuss
the question of whether the ANEC violations that they found were on
chronal or achronal geodesics.

With pure incoming states, they found that the ANEC integral vanished,
but at second order ANEC could be violated.  In this case, the
geodesics in question are chronal.  Almost all first-order
perturbations obey the generic condition, and a complete null geodesic
satisfying the generic condition with ANEC integral zero will have
conjugate points.  Thus at first order, almost all geodesics are
chronal, and is not necessary to go to second order.  However, in the
case of mixed incoming states they found ANEC violations at first
order, and in this case we cannot be sure whether the geodesics are
chronal or not.

\section{Anomalous violation of ANEC}
\label{sec:anomaly}

Visser \cite{gr-qc/9409043}, expanding upon the added note in Ref.\
\cite{PHRVA.D44.403}, points out that the stress-energy tensor
has anomalous scaling.  If we make a scale transformation,
\be
g \to \bar g =\Omega^2 g
\ee
then
\be\label{eqn:scaling}
T^a_b(\bar g)=\Omega^{-4}\left(T^a_b(g) -8aZ^a_b\ln\Omega\right)
\ee
where $a$ is a constant depending on the type of field under
consideration, and
\be
Z^a_b =\left(\nabla_c\nabla^d+\frac{1}{2}R_c^d\right) {C^{ca}}_{db}
\ee
Thus if $\gamma$ is some geodesic with tangent $k^a$,
\be
\int_\gamma T^a_b(\bar g)k^a k_b
=\Omega^{-4}\left(T_\gamma -8a\ln\Omega J_\gamma\right)
\ee
where
\be
T_\gamma = \int_\gamma T^a_b(g)k^a k_b
\ee
is the original ANEC integral, and
\be
J_\gamma = \int_\gamma Z_{ab} k^a k^b\,.
\ee
Thus if $J_\gamma$ does not vanish, there will be a rescaled version
of this spacetime in which $J_\gamma$ dominates $T_\gamma$, so that
ANEC is violated.  However, the necessary rescaling is enormous.  For
example, for a scalar field, $a =1/(2880\pi^2)$.  Thus if the initial
$J_\gamma$ and $T_\gamma$ are of comparable magnitude, we will need
$\Omega$ of order $\exp(2880\pi^2)$.  If $J_\gamma < 0$, then the
rescaling is contraction, and the curvature radius will become far
less than the Planck length, so semiclassical analysis (including that
used to derive the expression for the anomaly in the first place) will
not be applicable.  If $J_\gamma > 0$, then the rescaling is dilation.
In that case, the curvature radius of the spacetime is increased by
$\Omega$, so the Einstein tensor $G^a_b$ is multiplied by
$\Omega^{-2}$, while the stress-energy tensor $T^a_b$ is multiplied by
$\Omega^{-4}$.  Thus $T^a_b$ is infinitesimal compared to $G^a_b$ and
so cannot contribute to a self-consistent spacetime with achronal
geodesics.  In either case, then, this phenomenon does not violate
self-consistent achronal ANEC as formulated above.

Alternatively, as pointed out in \cite{gr-qc/9409043},
one can implement the anomalous scaling by changing the
renormalization scale $\mu$.  However, the result of such a drastic
change in scale is a theory vastly different from general relativity,
since higher-order terms in the renormalized Lagrangian now enter with
large coefficients.  Such a situation is also far from the domain of
validity of the semiclassical approximation.

\section{Proofs using self-consistent achronal ANEC}
\label{sec:proofs}

Several theorems in general relativity have been proved using ANEC (or
some variation thereof) as a premise.  The proofs of these theorems
only require that ANEC hold on achronal geodesics, so they apply
equally when the premise is replaced by Condition 1.  In fact we can
rule out wormholes connecting different regions and time machine
construction using only Lemma 1.

\subsection{Topological censorship}

Topological censorship theorems state that no causal path can go
through any nontrivial topology.  They rule out such things as
traversable wormholes.  We use the formulation of Friedman, Schleich
and Witt \cite{gr-qc/9305017} with Condition 1 instead of regular
ANEC.  We must also restrict ourselves to simply-connected spacetimes,
which means that the wormholes we rule out are only those which
connect one asymptotically flat region to another, not those which
connect a region to itself.

\begin{theorem}[Topological censorship]
Let $M, g$ be a simply-connected, asymptotically flat, globally
hyperbolic spacetime satisfying Condition 1 and the generic
condition.  Then every causal curve from past null infinity
($\scri^-$) to future null infinity ($\scri^+$) can be deformed to a
curve near infinity.
\end{theorem}

Friedman and Higuchi \cite{F&H} (see also
\cite{gr-qc/9301015}) outline a simple proof of this theorem which
applies equally well in our context.  Suppose there is a causal curve
$\gamma$ from $\scri^-$ to $\scri^+$ that cannot be deformed to a
curve near infinity (because it goes through a wormhole, for example).
It is then possible to construct a ``fastest'' causal curve $\gamma'$
homotopic to $\gamma$, where one curve is (weakly) ``faster'' than
another if it arrives at $\scri+$ in the causal past and departs from
$\scri-$ in the causal future of the other.  Such a ``fastest'' causal
curve must be a complete null geodesic.  Since $M$ is simply
connected, if $\gamma'$ were chronal we could deform it to a timelike
curve, and then to a ``faster'' curve.  Thus $\gamma'$ is an
achronal, complete null geodesic, but such a geodesic is ruled out by
Lemma 1.

One can see the necessity of simple connectedness (or some other
additional assumption) by considering the following
example.\footnote{We thank Larry Ford for pointing out this
  counterexample.}  Let $M$ be a static spacetime with a single
asymptotically flat region and a wormhole connecting the region to
itself, and suppose the throat of the wormhole is longer than the
distance between the mouths on the outside.  Any causal path through
the wormhole emerges in the future of the place where it entered, and
thus is not achronal.  We can still find fastest paths through the
wormhole, but they are chronal.  This can happen because the timelike
connections between points on such a path are not in the same homotopy
class as the path itself.

\subsection{Closed timelike curves}

The first use of global techniques to rule out causality violation was
by Tipler \cite{Tipler:1976bi}.  His theorem and proof transfer
straightforwardly to self-consistent achronal ANEC.

\begin{theorem}[No construction of time machines --- Tipler version]
An asymptotically flat spacetime $M, g$ cannot be null geodesically
complete if (a) Condition 1 holds on $M, g$, (b) the
generic condition holds on $M, g$, (c) $M, g$ is partially
asymptotically predictable from a partial Cauchy surface $S$, and (d) the
chronology condition is violated in $J^+(S) \cap J^-(\scri^+)$.
\end{theorem}
In order for the chronology condition to be violated (i.e., in order
for there to be closed timelike curves), there must be a Cauchy
horizon $H^+(S)$, which is the boundary of the region $D^+(S)$ that is
predictable from conditions on $S$.  The Cauchy horizon is composed of
a set of null geodesic ``generators.''  Tipler \cite{Tipler:1976bi}
shows that conditions (c) and (d) imply that there is at least one
such generator $\eta$ which never leaves $H^+(S)$.  If the spacetime
were null geodesically complete, then $\eta$ would be a complete null
geodesic lying in $H^+(S)$.  No point of $H^+(S)$ could be in the
chronological future of any other such point, so $\eta$ would be a
complete, achronal null geodesic.  But Lemma 1 shows that no such
geodesic can exist if conditions (a) and (b) are satisfied.

A similar theorem was proved by Hawking \cite{Hawking:1991nk},
which we can similarly extend.

\begin{theorem}[No construction of time machines --- Hawking version]
Let $M, g$ be an asymptotically flat, globally hyperbolic spacetime
satisfying  self-consistent achronal ANEC and the generic condition,
with a partial Cauchy surface $S$.  Then $M,g$ cannot have a compactly
generated Cauchy horizon $H^+(S)$. 
\end{theorem}
The Cauchy horizon is compactly generated if the generators, followed
into the past, enter and remain within a compact set.  Hawking
\cite{Hawking:1991nk} shows that in such a case, there will be
generators which have no past or future endpoints.  As above, such generators
would be complete, achronal null geodesics, which cannot exist under
the given conditions.

\subsection{Positive mass theorems}

Penrose, Sorkin and Woolgar \cite{gr-qc/9301015} proved a positive
mass theorem based on ANEC.  Their proof depends only on the condition
that every complete null geodesic has conjugate points.  As they point
out, it is sufficient to require that every achronal, complete null
geodesic has conjugate points, and thus that there are no such
geodesics.

\subsection{Singularity theorems and superluminal communication}

Galloway \cite{Galloway:1981kk} and Roman
\cite{Roman:1986tp,Roman:1988vv} showed that a spacetime with a
closed trapped surface must contain a singularity if ANEC holds, but
the ANEC integral is taken not on a complete geodesic, but rather on a
``half geodesic'' originating on the surface and going into the
future.  The argument depends only on the fact that any such half
geodesic must have a point conjugate to the surface within finite
affine length.  But if the half geodesic were chronal, then it would
have such a conjugate point.  Thus a sufficient premise would be that
every achronal half geodesic must satisfy ANEC.

The problem with this ``half achronal ANEC'' condition is that it does
not hold for quantum fields, even in flat space.  A simple example is
a minimally-coupled scalar field in flat space with Dirichlet boundary
conditions in the $x$-$y$ plane.  Consider a null geodesic in the
positive $z$ direction starting at some $z = z_0 > 0$.  On this
geodesic, $T_{ab}k^a k^b = - 1/(16\pi^2 z^4)$, so the half ANEC
integral can be made arbitrarily negative by making $z_0$ small.
While this system is not self-consistent (nor does it obey the generic
condition), it is hard to imagine that a self-consistent version could
not be created, for example using a domain wall \cite{hep-th/0211244}.
Thus our weakened version of ANEC is just as effective as the standard
one, but in either case it is necessary to add additional
qualifications to the singularity theorems in order for them to be
obeyed by quantum fields.

No-superluminal-communication theorems are similar to singularity
theorems.  Ref.\ \cite{gr-qc/9805003} defined a superluminal travel
arrangement as a situation in which a central null geodesic leaving a
flat surface arrives at a destination flat surface earlier than any
other null geodesic, and proved that such a situation requires weak
energy condition violation.  The argument is that the null geodesics
orthogonal to the surface are parallel when emitted, but diverge at
the destination surface, and thus must be defocused.  Such defocussing
means that ANEC must be violated, with the integral along the path
from the source to the destination.

Since a chronal geodesic could not be the fastest causal path from one
point to another, it is sufficient to require that ANEC hold on
achronal partial geodesics.  But once again, this principle is easily
violated.  An example using the Casimir effect is discussed in Ref.\
\cite{gr-qc/9805003}.  So, as with singularity theorems,
self-consistent achronal ANEC is an adequate substitute for ordinary
ANEC, but additional constraints are necessary to rule out
superluminal communication.

\section{Discussion}

A longstanding open question in general relativity is what principle
--- if any --- prevents exotic phenomena such as time travel.
Standard energy conditions on the stress-energy tensor, such as
ordinary ANEC, provide well-motivated means for restricting exotic
phenomena, but suffer from known violations by simple quantum systems.
We have discussed here an improved energy condition, self-consistent
achronal ANEC.  It is strong enough to rule out exotic phenomena as
effectively as ordinary ANEC, but weak enough to avoid known
violations.  The key qualification is the restriction to achronal
geodesics, which both disallows several known violations of ordinary
ANEC and is a necessary condition to apply techniques that have been
used to prove ANEC for models in flat space.

\section{Acknowledgments}

N.\ G.\ was supported by National Science Foundation (NSF) grant
PHY-0555338, by a Cottrell College Science Award from Research
Corporation, and by Middlebury College.  K.\ D.\ O.\ was supported in
part by grant RFP1-06-024 from The Foundational Questions Institute
(fqxi.org).  We thank Eanna Flanagan, Larry Ford, Tom Roman and Matt
Visser for helpful conversations.

\bibliography{no-slac,paper}

\end{document}